\documentclass[aps,prc,twocolumn,amsmath,showpacs, superscriptaddress,floatfix,nofootinbib]{revtex4-1}

\usepackage{setspace,ulem, epsfig,amssymb,amsfonts,amsmath,mathtools,bm,color,xcolor,graphicx,braket,adjustbox,esint,upgreek}
\usepackage{ctable}

\newcommand{\Bc}  {B_c^+}

\begin{document}
%\begin{spacing}{1.0}
\title{$B_c^+$ formation from random charm and anti-bottom quarks in the quark-gluon plasma}
\author{Baoyi Chen}\email{baoyi.chen@tju.edu.cn}
\affiliation{Department of Physics, Tianjin University, Tianjin 300350, China}
\author{Liuyuan Wen} 
\affiliation{Department of Physics, Tianjin University, Tianjin 300350, China}
\author{Yunpeng Liu}
\affiliation{Department of Physics, Tianjin University, Tianjin 300350, China}
\date{\today}

\begin{abstract}
We study the $B_c^+$ production in Pb-Pb collisions at $\sqrt{s_{NN}}=5.02$ TeV. In the 
quark-gluon plasma (QGP) produced in heavy-ion collisions, 
heavy quarks make random motions with the energy loss. 
We employ the Langevin equations to study the non-equilibrium distributions 
of heavy quarks and 
the Instantaneous Coalescence Model (ICM) to study the hadronization process. 
Due to abundant charm and bottom quarks in the QGP, their coalescence probability 
is significantly enhanced compared with the situations in proton-proton collisions.  
We find that the final production of $B_c^+$ 
is increased by the coalescence process, which makes 
the nuclear modification factor ($R_{AA}$) 
of $B_c^+$ larger than unit. 
Our model explains the experimental data 
well at semi-central and central collisions. The observation of $R_{AA}(B_c^+)>1$ is 
regarded as an evident and strong signal of the existence of the deconfined medium 
generated in heavy-ion collisions.

\end{abstract}
\pacs{14.40.Ld,14.40.Nd, 12.38.Mh, 25.75.-q}
\maketitle
%\section{Introduction}
The extremely hot QCD matter called ``quark-gluon plasma'' (QGP) is believed 
to be generated in the relativistic heavy-ion collisions~\cite{Bazavov:2011nk,NA50:1996lag}. 
The abnormal suppression of $J/\psi$ production in nucleus-nucleus (AA) has been 
regarded as one of the clean signals of this deconfined matter~\cite{Matsui:1986dk}. 
In the QGP, primordially 
produced $J/\psi$ in the hadronic collisions 
are dissociated by the color screening effect
and the inelastic scatterings with thermal 
partons~\cite{Satz:2005hx,
Yan:2006ve,Zhu:2004nw,Shi:2017qep,Zhao:2011cv,Du:2018wsj,Yao:2020xzw,Yao:2020eqy}. 
Both of these effects reduce the final production of $J/\psi$, 
which make the nuclear modification factor ($R_{AA}$) 
of $J/\psi$ smaller than 1. $R_{AA}$ 
is defined as the ratio between $J/\psi$ production 
in AA collisions and the yield in proton-proton (pp) collisions scaled with the 
number of binary collisions $N_{coll}$. At the Relativistic Heavy-Ion Collider (RHIC) and 
the Large Hadron Collider (LHC), multiple heavy-quark pairs are produced in parton 
hard scatterings. They combine to form new heavy quarkonium states at 
the hadronization of the medium~\cite{Andronic:2006ky,Blaizot:2017ypk,Blaizot:2018oev,
Chen:2017duy,Zhao:2017yan}. 
This new process is called coalescence, which can increase 
the nuclear modification factor of $J/\psi$ at the LHC collision 
energies~\cite{Thews:2000rj,Braun-Munzinger:2000csl, Zhao:2020jqu}. 

For charmonium and bottomonium, their nuclear modification factors are usually smaller 
than unit~\cite{STAR:2009irl,ALICE:2016flj,Du:2017qkv,Brambilla:2020qwo}. 
Recently, experiments have measured the $\Bc$ nuclear modification factors 
at $\sqrt{s_{NN}}=5.02$ TeV Pb-Pb collisions~\cite{ref-Bc-exp}. Its value is larger than 
the unit. This is a new strong and clear signal 
of the existence of QGP. 
{ In the coalescence process, 
the $B_c^+$ nuclear modification factor is proportional to the 
number of binary collisions, ${N_{b\bar b}N_{c\bar c}\over N_{\Bc}}\propto N_{coll}$.  }
The phenomenon of $R_{AA}(\Bc)>1$ is explained with the coalescence of 
random $\bar b$ and $ c$ quarks in QGP.  
This coalescence probability will be significantly enhanced when  
the numbers of charm and bottom pairs become large in the medium. 
As heavy quarks, especially bottom quarks,  
are hard to reach kinetic equilibrium due to their large masses, we will employ the 
Langevin equations to study the realistic dynamical evolutions of heavy quarks 
in the QGP and the Instantaneous Coalescence Model (ICM) to study the reaction of 
$\bar b+ c\rightarrow \Bc +g$ (where $g$ represents a gluon)~\cite{Cao:2015hia,
He:2011qa,Chen:2016mhl,Chen:2021akx}.  
Considering both the cold nuclear matter effect (such as the shadowing effect) and 
hot medium effects on heavy quarks and $\Bc$ mesons, we calculate the 
final spectrum of $\Bc$ mesons in Pb-Pb collisions 
and compare with the recent experimental data.

%\section{$\Bc$ in-medium properties}

In vacuum, heavy quark potential is parametrized with the Cornell potential. 
The binding energies and the wave functions of the eigenstates can be obtained 
by solving time-independent Schr\"odinger equation $H(r)R_{nl}(r)=E_{nl}R_{nl}(r)$, 
where $R_{nl}$ is the radial part of the total wave function. The angular part of the 
wave function is taken as spherical harmonics $Y_{lm}(\theta, \phi)$, where $(n,l,m)$ are the 
quantum numbers of the eigenstates. The form of Cornell potential is
\begin{align}
 V_{\mathrm{Cornell}}(r)=-{\alpha\over r}+\sigma r , 
\label{fun-cornell}
\end{align}
where the parameters are   
$\alpha={\pi/12}$ and $\sigma= 0.2\ \mathrm{GeV^2}$~\cite{Satz:2005hx}.

\begin{figure}[!hbt]
\centering
\includegraphics[width=0.3\textwidth]{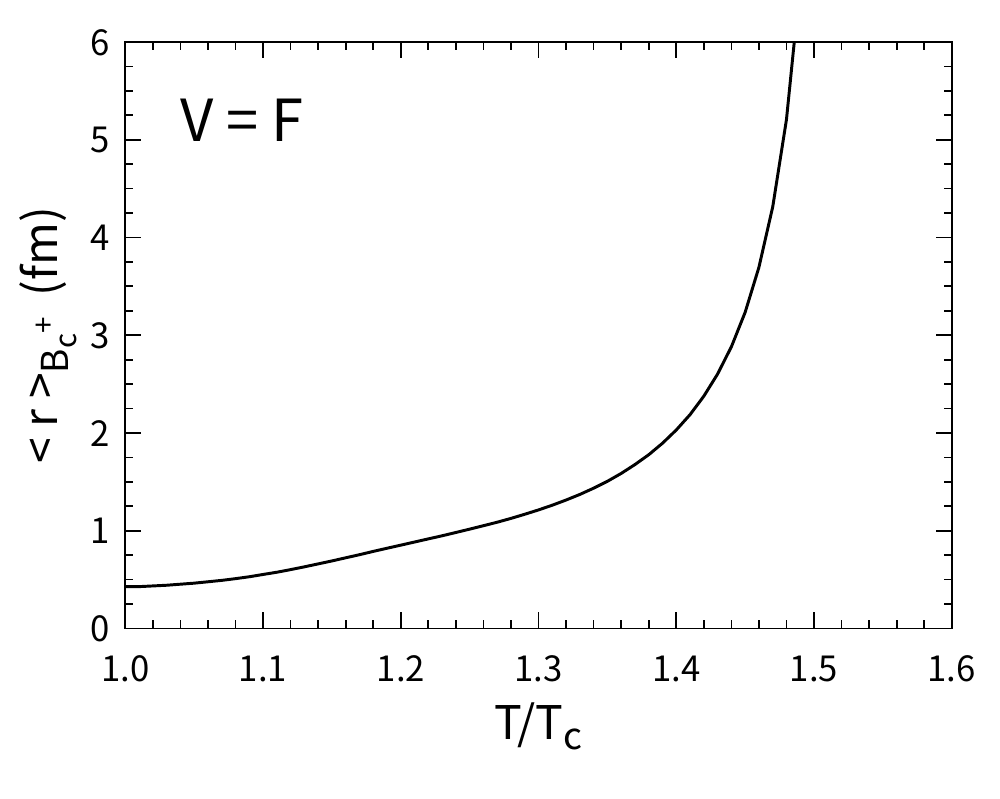}
\caption{Mean radius of $\Bc$ ground state as a function of temperature. The heavy quark 
potential is taken as the free energy $V=F(T,r)$. $T_c$ is the critical temperature 
of the phase transition.  
}
\hspace{-0.1mm}
\label{lab-figr}
\end{figure}

In the hot dense medium, both the coulomb term and the linear term in the Cornell potential are 
screened by the thermal partons due to the color screening effect. This screening 
effect becomes more vital at the larger distance and the higher temperature. 
{ 
In-medium heavy quark potential has been studied by complex-valued potential model~\cite{Lafferty:2019jpr} and lattice QCD calculations~\cite{Kaczmarek:2002mc,Kaczmarek:2005ui}. 
The realistic potential is between two limits: the free energy $F$ and the internal energy $U$. In this work, we don't intend to determine the exact formula of in-medium heavy quark potential. Instead, the potential is used to estimate the parameters in the Wigner functions which will be used in the coalescence process of $B_c^+$. }
The color screened potential is parametrized with the formula~\cite{Kaczmarek:2005ui},
\begin{align}
\label{fun-latreal}
F(T,r) =& -{\alpha\over r}(e^{-\mu r}+\mu r) 
-{\sigma \over 2^{3/4}\Gamma[3/4]}({r\over \mu})^{1/2} K_{1/4}[(\mu r)^2] \nonumber \\
&+ {\sigma 
\over 2^{3/2}\mu }{\Gamma[1/4]\over \Gamma[3/4]}
\end{align}
where $\alpha$ and $\sigma$ are the same as in the Cornell potential 
Eq.(\ref{fun-cornell}). 
The $\Gamma$ and $K_{1/4}$ are the Gamma function and the modified Bessel 
function respectively. 
The screening mass 
$\mu\equiv\mu({\bar T})$ from Lattice results is parametrized as, 
\begin{align}
{\mu(\bar T)\over \sqrt{\sigma}} = s\bar{T} +a \sigma_t \sqrt{\pi \over 2} 
[\mathrm{erf}({b\over \sqrt{2}\sigma_t}) - \mathrm{erf}({b-\bar{T}\over \sqrt{2}\sigma_t})]
\end{align}
where ${\bar T}\equiv T/T_c$, and s=0.587, a=2.150, b=1.054, $\sigma_t=0.07379$. 
$\mathrm{erf}(z)$ is the error function. 
With the increase of the temperature, 
heavy quark potential inside quarkonium is 
screened. { 
The mean radius of bound states 
increases with $T$, and approach to 
infinity at a certain temperature where 
the bound state is totally screened~\cite{Liu:2012tn}. In Fig.\ref{lab-figr}, we plot the mean radius $\langle r\rangle_{\Bc}(T)$ of $\Bc$ 
at different temperatures when taking the potential to be $V=F(r,T)$. 
At temperatures around the critical phase transition, the mean radius of $B_c^+$ is 
around 0.4 fm and increase to $\sim0.8$ fm at $1.2T_c$. 
The exact values of $B_c^+$ 
in-medium geometry size depend on the choice of in-medium potentials.  
If we approximate the $B_c^+$ wave function 
as a Gaussian function in a simple situation, 
the width of the Wigner function is determined with the $B_c^+$ geometry size. 
In order to study the effects of different in-medium potentials on $B_c^+$ production, 
different values of the width in the Wigner function will be considered. 
}

%\section{Langevin-Coalesence model}
%\subsection{Random evolutions of heavy quarks}
When heavy quarks move in the QGP, they lose energy via quasi-elastic scatterings 
with thermal light partons and the gluon radiation due to multiple scatterings. 
The random motion of heavy quarks is treated as Brownian motions. 
Langevin equations have been widely applied to simulate the time 
evolutions of the momentum distributions of heavy quarks. The classical Langevin equation 
with the gluon radiation is written as, 
 \begin{align}
\label{lan-gluon}
{d{\bf p}\over dt}= -\eta_D(p) {\bf p} +{\bf \xi} + {\bf f_g}
\end{align}
where ${\bf p}$ is the momentum of heavy quarks. { 
Neglect the momentum dependence in  
the white noise}, ${\bf \xi}$ is determined with the relation, 
\begin{align}
\langle \xi^{i}(t)\xi^{j}(t^\prime)\rangle =\kappa \delta ^{ij}\delta(t-t^\prime)
\end{align}
$i$ and $j$ 
represent three dimensions. $t$ is the time of heavy quark evolutions. 
$\kappa$ is the 
momentum diffusion coefficient of heavy quarks, which does not depend on 
the momentum of heavy quarks~\cite{Moore:2004tg}. 
It is connected with the spatial diffusion coefficient 
$\mathcal{D}_s$ with the relation $\kappa = 2T^2/\mathcal{D}_s$. The value of $\mathcal{D}_s$ 
is around $\mathcal{D}_s (2\pi T)\approx 5$ in the QGP~\cite{He:2012df}. 
The drag term 
$\eta_D(p)$ is determined with the fluctuation-dissipation relation,  
$\eta_D(p)=\kappa/( 2TE)$. Here $E= \sqrt{m^2+|{\bf p}|^2}$ is the 
energy of heavy quarks. 

The medium-induced gluon radiation term ${\bf f_g}= -{d{\bf p_g}/dt}$ 
contributes the recoil force on heavy quarks when 
they move through QGP~\cite{Zhang:2003wk}. 
${\bf p_g}$ is the momentum of emitted gluon. The 
number of radiated gluons in the time interval $t\sim t+\Delta t$ is, 
\begin{align}
\label{gluon-spec}
P_{\rm rad}(t,\Delta t) = \langle N_g(t, \Delta t)\rangle = \Delta t \int dx d k_T^2
{dN_g\over dx dk_T^2dt}
\end{align}
when the time interval $\Delta t$ is sufficiently small, the 
radiated gluons become smaller than the unit. 
Then $P_{\rm rad}$ can be interpreted as the probability 
to radiate a gluon in this time interval. 
$x=E_g/E$ is the ratio of the emitted gluon energy and 
the heavy quark energy. $k_T$ is the transverse momentum of the radiated gluon. 
$dN_g/dxdk_T^2dt$ is the spectrum of emitted gluons per unit time. 
{ It is adopted from the higher-twist calculation for the
medium-induced gluon radiation in perturbative QCD, 
where the spectrum of gluons radiated from massive heavy 
quarks is introduced by Ref.\cite{Zhang:2003wk}. }
Gluon radiation term dominates the energy loss of heavy quarks at high $p_T$ bins. 
{ At low 
$p_T$ bins, the contribution of gluon radiation becomes small, and 
the drag and random terms dominate the energy loss of heavy quarks. 
Relative magnitudes of two energy loss processes are compared in Ref.\cite{Cao:2015hia}.
Both collision process and 
the radiative process have been considered in the below figures. }

Bottom and charm pairs are produced in parton hard scatterings 
at the beginning of nucleus-nucleus collisions. Their density is proportional to the 
number of binary collisions in the overlap area of two nuclei. 
The initial spatial density of heavy quarks in Pb-Pb collisions 
is proportional to the 
thickness functions, 
\begin{align}
\label{eq-initqq-x}
{dN^{Q\bar Q}_{PbPb}\over d{\bf x_T}} = \sigma_{pp}^{Q\bar Q}
\times T_A({\bf x_T}-{\bf b}/2) T_B({\bf x_T}+{\bf b}/2) \mathcal{R}_s({\bf x_T}). 
\end{align} 
$T_{A(B)}({\bf x_T})=\int dz \rho({\bf x_T}, z)$ is the 
thickness function of nucleus A(B), which is defined as 
the integration of the nucleon density $\rho$ 
over the longitudinal coordinate $z$. 
${\bf b}$ is the impact parameter. It characterize 
the distance between the centers of two nuclei in the transverse plane. 
At 5.02 TeV pp collisions, 
the production cross section of charm quarks in the central rapidity ($|y|<2.3$) 
has been measured by ALICE 
Collaboration, ${d\sigma_{pp}^{c\bar c}/dy}=1.165$ mb~\cite{ALICE:2021dhb}. 
Bottom production cross section at 5.02 TeV is extracted with the pp collision data 
at 1.96 TeV and 2.76 TeV. The differential cross section 
is fitted to be $d\sigma_{pp}^{b\bar b}/dy = 47.5\ \mu b$ in the central rapidity 
($|y|<2.3$)~\cite{Chen:2017duy}. 
The initial momentum distributions of charm and bottom quarks are calculated 
with FONLL model~\cite{ref-FONLL}. 
In the central rapidity of Pb-Pb collisions at 5.02 TeV, 
shadowing effect $\mathcal{R}_s({\bf x_T})$ reduces the number of charm pairs by around 
25\%. We use the EPS09 model to generate the shadowing factor 
in our calculations~\cite{Eskola:2009uj}, which 
modifies the initial spatial and 
momentum distributions of charm pairs in Pb-Pb collisions. 
{ 
The shadowing effect in bottom quark distribution becomes weaker due to the larger mass. 
It is also included in the same way as charm quarks. }

In order to calculate the nuclear modification factor ($R_{AA}$) of $\Bc$, 
we also need the $\Bc$ production cross section in pp collision, 
which is used in the denominator of $R_{AA}$, 
\begin{align}
R_{AA} = {N_{AA}^{\Bc}\over N_{coll}N_{pp}^{\Bc}}. 
\end{align}
The numerator $N_{AA}^{\Bc}$  
is the production of $\Bc$ from the coalescence of random charm and anti-bottom quarks 
in Pb-Pb collisions. In the denominator,   
$N_{pp}^{\Bc}$ is the production of $\Bc$ in pp collisions, which is proportional to the 
production cross section $\sigma_{pp}^{\Bc}$. 
This cross section has been only measured at 1.96 TeV pp collisions 
by CDF Collaboration~\cite{CDF:2016hra}. 
At $p_T>6 $ GeV/c and the central rapidity $|y|<1$, the differential 
cross section of $\Bc$ with the branching ratio is 
$\sigma(\Bc)\mathcal{B}(\Bc\rightarrow J/\psi \mu^+\nu)= 0.60\pm 0.09$ nb. In order to 
parametrize the values of $\sigma(\Bc)$ at 5.02 TeV pp collisions, 
we firstly employ PYTHIA to calculate the 
values of $\sigma(\Bc)$ at 
1.96 TeV and 5.02 TeV pp collisions respectively, and obtain the ratio 
$\sigma(\Bc,5.02\mathrm{TeV})/\sigma(\Bc, 1.96\mathrm{TeV})=2.40$. 
Then $\Bc$ production cross section at 5.02 TeV is extracted to be 
$\sigma(\Bc)\mathcal{B}(\Bc\rightarrow J/\psi \mu^+\nu)= 2.4\times 0.6$ nb 
by scaling with the ratio. The branching ratio $\mathcal{B}(\Bc\rightarrow J/\psi \mu^+\nu)$ 
has been calculated with different theoretical models. 
The value of the branching ratio is predicted to be  
$\mathcal{B}(\Bc\rightarrow J/\psi \mu^+\nu)=(1.15-2.37)\%$ 
by QCD sum rules~\cite{Huang:2008zg}, 
(non-)relativistic constituent-quark models~\cite{Ebert:2003cn,Hernandez:2006gt}, 
QCD relativistic-potential models~\cite{AbdEl-Hady:1999jux} respectively. 
While nonrelativistic QCD model 
predicts the value to be $6.7^{+2.5}_{-1.4}\%$~\cite{Qiao:2012vt}. 
This contributes large uncertainty 
in the determination of $d\sigma_{pp}^{\Bc}/dy$. In this work, we take the branching ratio 
to be $2.37\%$ and the middle value of two scenarios $4.54\%$. The 
production cross section of $\Bc$ at 5.02 TeV is 
${d\sigma_{pp}^{\Bc}\over dy}|_{p_T>6\mathrm{GeV/c}}= 30.38$ nb (taking 
$\mathcal{B}=2.37\%$) and $15.86$ nb (taking 
$\mathcal{B}=4.54\%$) respectively in the central rapidity 
$|y|<2$. This production cross section has included the contributions from 
the decay of $\Bc$ excited states. 

In order to obtain the $\Bc$ production cross section at other $p_T$ bins, we use 
PYTHIA to generate the normalized transverse momentum distribution, 
\begin{align}
{dN_{pp}^{\Bc}\over 2\pi p_T dp_T} ={(n-1)\over \pi (n-2)\langle p_T^2\rangle_{pp}} 
[1+{p_T^2\over (n-2)\langle p_T^2\rangle_{pp}}]^{-n} 
\end{align}
where the mean transverse momentum square of $\Bc$ in the central rapidity 
is $\langle p_T^2\rangle_{pp}=25.1\ \mathrm{(GeV/c)^2}$. The 
value of $n$ is $n=4.16$. With this normalized 
distribution, we extract 
the production cross section of $\Bc$ at other $p_T$ bins by scaling with the 
values of $d\sigma_{pp}^{\Bc}/dy$ at $p_T>6$ GeV/c. We obtain the relation 
${d\sigma_{pp}^{\Bc}\over dy}|_{p_T>11 \mathrm{GeV/c}} = 3.65$ nb 
(taking $\mathcal{B}=2.37\%$) and $1.90$ nb (taking $\mathcal{B}=4.54\%$) respectively.   
Furthermore, the total cross section of $\Bc$ in $p_T>0$ becomes 
${d\sigma_{pp}^{\Bc}\over dy}|_{p_T>0 } = 5.0\times
{d\sigma_{pp}^{\Bc}\over dy}|_{p_T>6\mathrm{GeV/c}}$.

In Pb-Pb collisions, hot deconfined matter consisting of light quarks and gluons 
turns out to be a strong coupling matter. 
Its dynamical evolutions are described with hydrodynamic equations. We use the 
well developed MUSIC package to simulate the time evolutions of local temperatures 
and velocities of the expanding QGP~\cite{Schenke:2010rr}. 
Equation of state (EoS) is needed to close the 
hydrodynamic equations. 
In the deconfined phase where the medium temperature is above 
the critical temperature $T_c=170$ MeV, 
EoS is taken from Lattice QCD calculations. 
In the hadronic phase, EoS is taken from the 
Hadron Resonance Gas model (HRG)~\cite{Huovinen:2009yb}. 
Two phases are 
connected with the crossover phase transition. 
The initial temperature of the 
medium can be determined by the final multiplicity of light hadrons measured in experiments. 
The initial temperature of QGP is fitted as $T(\tau_0, {\bf x_T}=0)=510$ MeV in the 
most central collisions with the impact parameter ${\bf b}=0$. 
$\tau_0=0.6$ fm/c is the time 
of QGP reaching local equilibrium. The value of $\tau_0$ is determined 
by fitting the anisotropic flows of light hadrons measured in experiments. 
Hydrodynamic equations start evolutions from the time $\tau=\tau_0$. 
Evolutions of heavy flavors in the pre-equilibrium stage have been 
neglected.

%\subsection{Hadronization process}
In the QGP, when the local 
temperature of QGP is higher than the dissociation temperature of $\Bc$, 
charm and anti-bottom quarks do not form a bound state, as the 
thermal medium screens their interaction. 
Instead, they make Brownian motions in the medium independently. 
When heavy quarks move to the regions where the 
local temperature of QGP is smaller than a certain value, heavy quark potential is 
partially restored. These heavy quarks have a probability to 
hadronize into a bound 
state. As the binding energies of $\Bc$ states are larger than the values of light hadrons, 
$\Bc$ states can survive at $T>T_c$. 
{
We take the coalescence temperature of $\Bc$ 
to be $T_{\rm coal}=1.2\ T_c$, where $\Bc$ wave function at 
this temperature is close to the situation in vacuum, as shown in Fig.\ref{lab-figr}. 
If the coalescence process happens at a higher temperature where 
$\Bc$ wave function is significantly modified by the medium, Landau damping effect will 
dissociate most of the $\Bc$ generated at the high temperatures. Only those 
$\Bc$ generated at relatively low temperatures can survive in the QGP. 
The coalescence probability 
is given by the Wigner function $f^W({\bf x_r},{\bf q_r})$ of $\Bc$ mesons, which 
is connected with $\Bc$ wave function via Weyl transform. 
Instead of employing realistic in-medium wave functions of $B_c^+$, we approximate the 
wave function to be a Gaussian function and get the corresponding Wigner function, 
$f^W({\bf x_r}, {\bf q_r})
= 8\exp[-{{x_r}^{2}\over \sigma^2} - \sigma^2 {q_r}^2]$. 
${\bf x_r} \equiv {\bf x}_1^{\rm cm} - {\bf x}_2^{\rm cm}$ and $
{\bf q_r} \equiv {E_1^{\rm cm}{\bf p}_1^{\rm cm} 
- E_2^{\rm cm} {\bf p}_2^{\rm cm} \over E_1^{\rm cm} + E_2^{\rm cm}}$ 
are the relative position and momentum between two quarks 
in the center of mass frame (CoM). ${\bf x}_{1,2}^{\rm cm}$ 
and ${\bf p}_{1,2}^{\rm cm}$ are the positions 
and momenta of two particles in the CoM-frame. 
$E_{1,2}^{\rm cm}=\sqrt{m_c^2+|{\bf p}_{1,2}^{\rm cm}|^2}$ is the energy of the
heavy (anti)quark.
The width of the Wigner function 
is determined with the root-mean-square radius of $\Bc$ via 
$\sigma^2= {4\over 3}{(m_1+m_2)^2\over m_1^2 +m_2^2}
\langle r^2\rangle_{\Bc} $~\cite{Chen:2021akx,Greco:2003vf}. 
$\langle r^2\rangle_{\Bc}$ can be calculated from the Schr\"odinger equation 
with in-medium potential. We treat it as a parameter in the model and take different values 
(e.g. $\sqrt{\langle r^2\rangle_{\Bc}}=0.3, 0.5, 1.0$ fm respectively) 
to consider the effects of different in-medium potentials on the $\Bc$ production. 
The mean coalescence probability of one charm and one anti-bottom quark in the hot medium 
is given by the Instantaneous Coalescence Model (ICM)~\cite{Chen:2021akx},
}
\begin{align}
\label{eq-psicoal}
&\langle \mathcal{P}_{\bar b c\rightarrow \Bc g}({\bf x_M}, {\bf p_M})\rangle \nonumber \\
&=
g_d \int d{\bf x_1} d{\bf x_2}
{d{\bf p_1}\over (2\pi)^3} {d{\bf p_2}\over
(2\pi)^3} {d^2N_1\over d{\bf x_1}d{\bf p_1}}
{d^2N_2\over d{\bf x_2} d{\bf p_2}}
f^W({\bf x_r}, {\bf q_r})   \nonumber \\
&\qquad\qquad \qquad \times
\delta^{(3)}({\bf p_M} -{\bf p_1}-{\bf p_2})
\delta^{(3)}({\bf x_M} - {{\bf x_1} +{\bf x_2}\over 2}), 
\end{align}
where $\langle \mathcal{P}_{\bar b c\rightarrow \Bc g}\rangle$ is the ensemble averaged 
probability of one random $\bar b$ quark combining with one $c$ quark at the coalescence 
temperature in QGP.
{ Here $dN_{i}/d{\bf x}_{i}d{\bf p}_{i}$ is the phase space distribution of one particle.  ${\bf x_M}$ and ${\bf p_M}$ 
are the positions and momentum of the formed meson. 
${\bf x_{1,2}}$ and ${\bf p_{1,2}}$ are the positions and momentum of heavy (anti-)quarks 
respectively.} In the coalescence reaction  $\bar b+ c\rightarrow \Bc +g$, 
the momentum of the gluon has been neglected in the momentum conservation 
equation,   
${\bf p}_{\Bc}= {\bf p}_{\bar b}+{\bf p}_{ c}$, which is represented 
by the delta function in Eq.(\ref{eq-psicoal}). The
energy conservation in the reaction 
has been ignored in the present framework 
which will be included later~\cite{Yao:2017fuc,Blaizot:2021xqa}.
{Both $B_c^+$ and $B_c^+$(2s) are observed in experiments~\cite{ParticleDataGroup:2018ovx}. They are spin singlets, and it is natural to assume that their corresponding spin triplets exist. The radial excited states will decay into the ground state. We neglect the differences between all these states and count all of them together in $B_c^+$ production, resulting in $g_d=2/9$, where the factor $2$ comes from the radial excited states, and the factor $1/9$ comes from the color factor.}{}
{Charm and anti-bottom quarks are randomly generated according to initial 
spatial and momentum distributions given before. }
Their event-by-event 
stochastic evolutions in QGP are described with Langevin equations, 
and the Wigner function controls the coalescence 
probability. 
The yield 
of $\Bc$ in Pb-Pb collisions become, 
\begin{align}
\label{eq-hadron}
N^{B_c^+}_{AA}  =
\int d{\bf x_M} {d {\bf p_M} \over (2\pi)^3} 
\langle{\mathcal{P}_{\bar b c\rightarrow \Bc g}({\bf x_M},{\bf p_M})}\rangle
\times
{(N^{c\bar c}_{AA}  N^{b\bar b}_{AA})} ,
\end{align}
The final production of $\Bc$ is proportional to the number of 
charm and anti-bottom quarks in the QGP. The number of heavy quarks $N_{AA}^{Q\bar Q}$ in 
Pb-Pb collisions are obtained with Eq.(\ref{eq-initqq-x}). 

{ With above Langevin equation plus ICM, we can describe the 
non-equilibrium distributions
of heavy quarks in QGP, and their hadronization into hidden heavy flavor hadrons. 
This model has explained the experimental data of 
$J/\psi$ spectrum in 5.02 TeV Pb-Pb collisions~\cite{Chen:2021akx}. 
Due to the similarity between $J/\psi$ and $\Bc$ coalescence process, 
we apply this approach to the $\Bc$ study. }
%\section{$B_c^+$ production in heavy-ion collisions}
With event-by-event numerical 
simulations, we obtain the final 
spectrum of $\Bc$ produced in Pb-Pb collisions 
after considering both cold and hot nuclear 
matter effects.   
Now, we compare our theoretical results with the experimental data recently 
measured by CMS Collaborations. 

\begin{figure}[!hbt]
\centering
\includegraphics[width=0.4\textwidth]{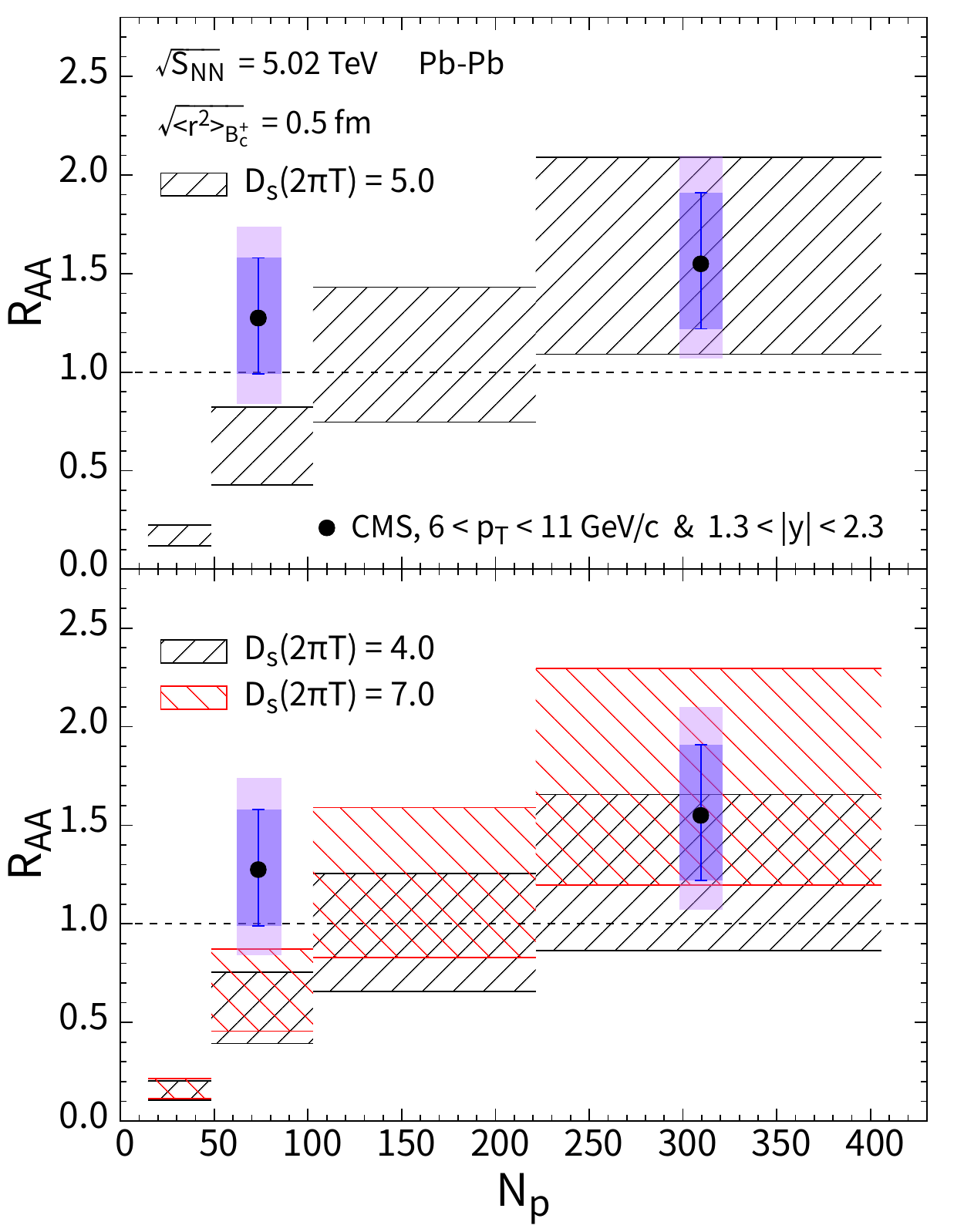}
\caption{$\Bc$ nuclear modification factor $R_{AA}$ as a function of the number 
of participants $N_p$ in the transverse momentum $6<p_T<11$ GeV/c 
in 5.02 TeV Pb-Pb collisions. Root-mean-square radius of $\Bc$ is taken as 
$\sqrt{\langle r^2\rangle_{\Bc}}=0.5$ fm in both upper and lower panel. 
The spatial diffusion coefficients satisfy the relation $\mathcal{D}_s(2\pi T)=5.0$ 
(in upper panel) and $\mathcal{D}_s(2\pi T)=4.0, 7.0$ (in lower panel), respectively. 
Theoretical bands correspond to the centrality bins: 
cent.0-20\%, 20-40\%, 40-60\%, 60-80\% respectively. 
The lower and upper limit of the theoretical bands correspond 
to a larger and smaller value of $d\sigma_{pp}^{\Bc}/dy$ respectively. Experimental 
data is from CMS Collaboration~\cite{ref-Bc-exp}. 
}
\hspace{-0.1mm}
\label{lab-RAANp}
\end{figure}

In Fig.\ref{lab-RAANp}, we calculate the $R_{AA}(\Bc)$ as a function of $N_p$ 
at $6<p_T<11$ GeV/c in the rapidity range $1.2<|y|<2.3$. In the central 
collisions, both theoretical and experimental results are above unit, which means 
that QGP can enhance the production of $\Bc$ compared with the production in pp 
collisions scaled with $N_{coll}$. This is due to the coalescence of random 
$\bar b$ and $ c$ quarks in QGP. 
The final production of $\Bc$ is proportional to the densities of charm and bottom 
quarks in the medium. 
With more heavy quarks in QGP, charm quarks have a larger probability to 
meet with another anti-bottom quark, before they hadronize into D mesons. 
As most of primordially produced $\Bc$ are dissociated by the QGP in central collisions, 
the final production 
of $\Bc$ is dominated by the coalescence process.  
The phenomenon of 
$R_{AA}>1$ in central and semi-central collisions indicate the existence 
of QGP and the coalescence mechanism. 
In peripheral collisions 
$N_p<100$, the temperatures of QGP become lower.  
As $\Bc$ can survive in the temperature 
region $T_c<T<T_d$ in QGP due to the large binding energy, when the temperature of QGP 
is slightly above $T_c$, some of primordially produced 
$\Bc$ can survive in the medium. 
Meanwhile, the production of $\Bc$ from the   
coalescence is reduced as less heavy quark pairs are produced in the peripheral collisions. 
Therefore, the theoretical results with only coalescence contribution are smaller than the 
experimental data at $N_p<100$ in Fig.\ref{lab-RAANp}. 
The band of the theoretical results is due to the uncertainty of 
$d\sigma_{pp}^{\Bc}/dy$ in the denominator of $R_{AA}$. 
{ In the upper panel of Fig.\ref{lab-RAANp}, the absolute yields of 
$\Bc$ in pp collisions scaled by $N_{coll}$ 
is ${dN_{pp}^{B_c^+}\over dy} N_{coll}=0.33-0.64$ in $6.0<p_T<11.0$ GeV/c in centrality 0-20\%. 
The $\Bc$ yields in Pb-Pb collisions can be obtained by scaling with the value of 
$R_{AA}$. 
In the lower panel of Fig.\ref{lab-RAANp}, 
different values of spatial diffusion coefficient are also employed to study the effects 
of heavy quark thermalization on $\Bc$ production. 
In the situation of 
$\mathcal{D}_s(2\pi T)=4.0$, bottom and 
charm quarks are strongly coupled with the hot medium. Their momentum 
distributions are more close 
to the equilibrium distribution, which reduces the $\Bc$ production in high $p_T$ bin (
such as in $6<p_T<11$ GeV/c) compared with the weak coupling 
situation of $\mathcal{D}_s(2\pi T)=7.0$. 
However, the total production of $\Bc$ in $p_T>0$ is enhanced by around 
$20\%$ in the case of $\mathcal{D}_s(2\pi T)=4.0$ than 
the case of $\mathcal{D}_s(2\pi T)=7.0$. 
Note that the values of $\mathcal{D}_s(2\pi T)$ 
in Fig.\ref{lab-RAANp} are still 
within the uncertainties of $\mathcal{D}_s$ determined by $D$ meson nuclear 
modification factors at RHIC and LHC collision energies~\cite{Cao:2015hia, Cao:2013ita}. 
}

%%%%%%%%%%%%%%%%%%%
\begin{figure}[!hbt]
\centering
\includegraphics[width=0.4\textwidth]{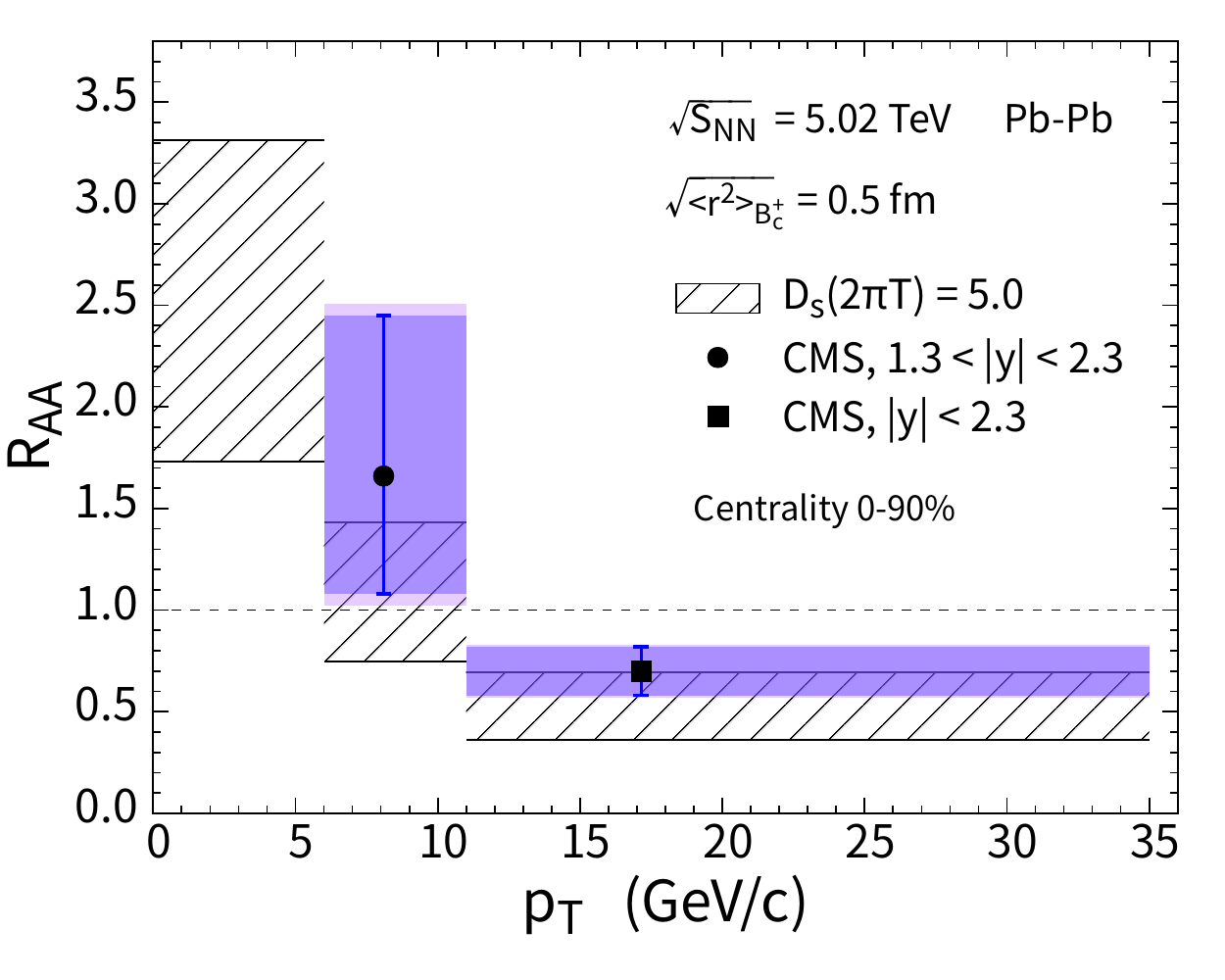}
\caption{ $p_T$ dependence of the nuclear modification factor 
$R_{AA}$ of $\Bc$ in centrality 0-90\% in 5.02 TeV Pb-Pb collisions. 
Root-mean-square radius of $\Bc$ is taken as $\sqrt{\langle r^2\rangle_{\Bc}}=0.5$ fm. 
Spatial diffusion coefficient is determined as $\mathcal{D}_s(2\pi T)=5.0$. 
Lower and upper limit of the 
theoretical results correspond to the situations with larger and smaller 
values of $d\sigma_{pp}^{\Bc}/dy$ respectively. Experimental data are 
from CMS Collaboration~\cite{ref-Bc-exp}.   
}
\hspace{-0.1mm}
\label{lab-RAApt}
\end{figure}
%%%%%%%%%%%%%

In Fig.\ref{lab-RAApt}, 
the $p_T$ spectrum of $\Bc$ in centrality 0-90\% is plotted.  
In the low $p_T$ region, the production of 
$\Bc$ in Pb-Pb collisions mainly come from the coalescence process. 
As heavy quarks are strongly coupled with the QGP, they lose energy when moving 
through the medium. Heavy quarks with large $p_T$ will be shifted to the 
moderate and low $p_T$ bins due to the energy loss. 
Then the 
$\Bc$ generated in the reaction $\bar b+c\rightarrow \Bc +g$ are mainly located 
in low and moderate $p_T$ bins. Therefore, $\Bc$ final production is dominated by the 
coalescence process in the low $p_T$ bins. This results in 
$R_{AA}>1$ at $p_T\lesssim 11$ GeV/c 
in Fig.\ref{lab-RAApt}. 
In the $11<p_T<35$ GeV/c, theoretical results 
is slightly smaller than the experimental data. This is because we do not include 
the primordial production which becomes more important in higher $p_T$ bins. 
The competition between the primordial production and the regeneration 
from coalescence process at different $p_T$ bins and centralities  
have already been observed in $R_{AA}$ of $J/\psi$.

{ In above calculations, one of important ingredients in coalescence process 
is the Wigner function. It is connected with the geometry size of $\Bc$ at the coalescence  
temperature. When taking a strong in-medium potential, the value of root-mean-square radius 
of $\Bc$ becomes smaller. 
In the situation of $\sqrt{\langle r^2\rangle_{\Bc}}=0.3$ fm, the $\Bc$ production 
is similar to the case of 0.5 fm in Fig.\ref{lab-RAANp} without evident difference. 
If the potential is very weak, we take the value of  
root-mean-square radius $\sqrt{\langle r^2\rangle_{\Bc}}=1.0$ fm. 
In the Wigner function, the probability in spatial coalescence condition is enhanced. 
However, the constraints on the relative momentum between heavy quarks 
become more strict. 
The combined effects
from spatial and momentum coalescence conditions consistently given by Wigner function
reduce the $\Bc$ production by around 50\% compared with the situation 
of  $\sqrt{\langle r^2\rangle_{\Bc}}=0.5$ fm. 
Another relevant 
parameter is the coalescence temperature of $\Bc$. Its value is taken as $1.2T_c$ according 
to previous discussions. With a smaller coalescence temperature such as $1.1T_c$, heavy quarks 
diffuse into a larger volume of the medium and their momentum distributions become 
slightly softer. $\Bc$ production is reduced by around $5\%$ in the case of lower coalescence 
temperature. 
}

%\section{Conclusion}
In this work, we employ the Langevin equation to simulate the time evolutions of 
heavy quarks in QGP, and obtain the  
non-equilibrium distributions of heavy quarks. 
The hadronization of $\Bc$ is studied via the ICM. With multiple charm and bottom pairs 
making independent Brownian motions in the hot medium, 
the coalescence probability of random 
charm and anti-bottom quarks are enhanced. From the theoretical calculations, in 
central collisions, 
most of the $\Bc$ final production is 
from the coalescence process 
instead of the primordial production. 
Furthermore, due to the significant energy loss of heavy quarks when they travel through 
the medium, the regenerated $\Bc$ are mainly located in the low and moderate $p_T$ bins. 
In the high $p_T$ bins, $\Bc$ production is dominated by the primordial production. 
Both experimental and theoretical studies show that 
the nuclear modification factor of $\Bc$ can be larger 
than 1.0 in Fig.\ref{lab-RAANp}-\ref{lab-RAApt}. 
%which is due to the coalescence of 
%heavy quarks produced in different hadronic collisions. 
The observation of $R_{AA}(\Bc)>1$ is 
regarded as an evident signal of the existence of the deconfined medium 
generated in Pb-Pb collisions.

\vspace{0.5cm}
{\bf Acknowledgement:}
This work is supported by the National Natural Science Foundation of China
(NSFC) under Grant Nos. 12175165, 11705125.

%%%%%%%%%%%%%%%%%%
%%%%%%%%%%%%%%%%%%
%\end{spacing}
\end{document}